\RequirePackage{etoolbox} %

\newtoggle{draftver}
\toggletrue{draftver}

\newtoggle{refereever}

\nottoggle{refereever}{
    \documentclass[bibyear]{aa}  
}{
    \documentclass[referee, bibyear]{aa}  
}

\usepackage[varg]{txfonts}

\usepackage[utf8]{inputenc}
\usepackage[T1]{fontenc}
\usepackage[english]{babel}

\usepackage{natbib}
\bibpunct{(}{)}{;}{a}{}{,} %
\usepackage{csquotes}
\nocite{*}

\let\cite\biblatexcite
\newcommand*{\cite}{\citealp}
\newcommand*{\textcite}{\citet}

\usepackage{amsmath}
\usepackage{amsfonts}
\usepackage{amssymb}
\usepackage{graphicx}

\usepackage{float}

\usepackage{diagbox}
\usepackage{booktabs}
\usepackage{multirow}
\usepackage{multicol}
\usepackage{makecell}
\usepackage{tabularx}
\usepackage{adjustbox}
\usepackage{placeins}[section]
\usepackage[separate-uncertainty=true, 
            separate-uncertainty-units = bracket,%
            range-phrase=-, 
            range-units=single]{siunitx}

\nottoggle{refereever}{
    \usepackage[table]{xcolor}
    
    \definecolor{lightgreen}{RGB}{144,238,144}
    \definecolor{lightcoral}{RGB}{240,128,128}
    \definecolor{darkgreen}{RGB}{0,100,0}

    \iftoggle{draftver}{
        \newcommand{\draftcomment}[1]{\textcolor{red}{#1}}
        \newcommand{\juan}[1]{\textcolor{blue}{#1}}
        \newcommand{\ivan}[1]{\textcolor{red}{#1}}
        \newcommand{\andrea}[1]{\textcolor{darkgreen}{#1}}
        \newcommand{\filippo}[1]{\textcolor{olive}{#1}}
    }{
        \newcommand{\draftcomment}[1]{}
        \newcommand{\juan}[1]{}
        \newcommand{\ivan}[1]{}
        \newcommand{\andrea}[1]{}
        \newcommand{\filippo}[1]{}
    }
}{}

\usepackage[
        hidelinks, 
        colorlinks=true,
        allcolors=blue,
        allbordercolors=blue,
        linktoc=all,
    ]{hyperref}

\usepackage{lscape} %
\usepackage{rotating} %

\usepackage{mathtools}

\title{The flaring activity of blazar \object{AO 0235+164} during year 2021.}

\date{Received 25 Feb 2024 / Accepted 15 May 2024}

\author{
	         Juan Escudero Pedrosa\thanks{\email{jescudero@iaa.es}}\inst{,\ref{affil:IAA-CSIC}}
	    \and Iván Agudo\inst{\ref{affil:IAA-CSIC}}
        \and Till Moritz \inst{\ref{affil:IAA-CSIC}}
	    \and Alan P. Marscher\inst{\ref{affil:BU-blazar}}
	    \and Svetlana Jorstad\inst{\ref{affil:BU-blazar},\ref{affil:SPBU}}
	    \and Andrea Tramacere\inst{\ref{affil:UniGe-astro}}
	    \and Carolina Casadio\inst{\ref{affil:UC-Greece},\ref{affil:IAFRT-Greece}}
	    \and Clemens Thum\inst{\ref{affil:IRAM-Granada}}
	    \and Ioannis Myserlis\inst{\ref{affil:IRAM-Granada}}
	    \and Albrecht Sievers \inst{\ref{affil:IRAM-Granada}}
	    \and Jorge Otero-Santos\inst{\ref{affil:IAA-CSIC}}
	    \and Daniel Morcuende\inst{\ref{affil:IAA-CSIC}}
	    \and Rubén López-Coto\inst{\ref{affil:IAA-CSIC}}
	    \and Filippo D'Ammando\inst{\ref{affil:INAF-ira}}
	    \and Giacomo Bonnoli\inst{\ref{affil:INAF-brera},\ref{affil:IAA-CSIC}}
        \and Mark Gurwell\inst{\ref{affil:SMA}}
    \and José Luis Gómez\inst{\ref{affil:IAA-CSIC}}
    \and Ramprasad Rao\inst{\ref{affil:SMA}}
    \and Garrett Keating\inst{\ref{affil:SMA}}
}
\institute{
	         Instituto de Astrofísica de Andalucía, CSIC, Glorieta de la Astronomía s/n, 18080 Granada\label{affil:IAA-CSIC}
	    \and Institute for Astrophysical Research, Boston University, 725 Commonwealth Avenue, Boston, MA 02215, United States\label{affil:BU-blazar}
        \and Department of Astronomy, University of Geneva, ch. d’Ecogia 16, 1290 Versoix, Switzerland\label{affil:UniGe-astro}
	    \and Department of Physics, University of Crete, 71003, Heraklion, Greece\label{affil:UC-Greece}
	    \and Institute of Astrophysics, Foundation for Research and Technology - Hellas, Voutes, 70013 Heraklion, Greece\label{affil:IAFRT-Greece}
	    \and Institut de Radioastronomie Millimétrique, Avenida Divina Pastora 7, Local 20, E-18012, Granada, Spain\label{affil:IRAM-Granada}
	    \and INAF - Istituto di Radioastronomia, Via Gobetti 101, I-40129 Bologna, Italy\label{affil:INAF-ira}
	    \and INAF Osservatorio Astronomico di Brera, Via E. Bianchi 46, 23807 Merate (LC), Italy\label{affil:INAF-brera}
	    \and Saint Petersburg State University, 7/9 Universitetskaya nab., St. Petersburg, 199034 Russia\label{affil:SPBU}
        \and Center for Astrophysics — Harvard \& Smithsonian, 60 Garden Street, Cambridge, MA 02138 USA\label{affil:SMA}
}

\begin{document}

        \abstract{}{}{}{}{}
          \abstract
           { The blazar \object{AO 0235+164}, located at redshift $z=0.94$, has displayed interesting and repeating flaring activity in the past, the latest episodes occurring in 2008 and 2015. In 2020, the source brightened again, starting a new flaring episode that peaked in 2021.}
           { We study the origin and properties of the 2021 flare in relation to previous studies and the historical behavior of the source, in particular to the 2008 and 2015 flaring episodes. }
           { We analyze the multi-wavelength photo-polarimetric evolution of the source. From Very Long Baseline Array images, we derive the kinematic parameters of new components associated with the 2021 flare. We use this information to constrain a model for the spectral energy distribution of the emission during the flaring period. We propose an analytical geometric model to test whether the observed wobbling of the jet is consistent with precession.}
           {We report the appearance of two new components that are ejected in a different direction than previously, confirming the wobbling of the jet. We find that the direction of ejection is consistent with that of a precessing jet.
           The derived period independently agrees with the values commonly found in the  literature.
            Modeling of the spectral energy distribution further confirm that the differences between flares can be attributed to geometrical effects.}
           {}
   
        \keywords{Astroparticle physics -- Accretion, accretion disks -- Polarization -- Radiation mechanisms: general --  Galaxies: jets -- Relativistic processes}
        
        \maketitle

    \section{Introduction}
    
    Blazars, a type of active galactic nuclei (AGNs), are amongst the most energetic objects in the Universe. They are generally accepted to consist of a super-massive black hole (BH) surrounded by an accretion disk and usually a dusty torus (DT), with symmetrical jets of matter emanating from the vicinity of the black hole that can extend far beyond the size of its host galaxy. The exact mechanisms by which high energy emission from blazars is generated is not well understood, and questions remain about the exact mechanisms by which plasma in the jet is collimated and accelerated to speeds close to that of light, as well about the particle composition of the jet and the location and cause of the observed variability and $\gamma$-ray emission. 
    
    \object{AO 0235+164} is a BL Lacertae-type blazar located at redshift $z = 0.94$ (\cite{Cohen:1987}). It is known to exhibit strong variability across the entire spectrum, and has repeatedly displayed high-amplitude flaring behavior in recent years. In particular, episodes in 2008 (\cite{Agudo:2011}) and 2015 (\cite{juan_0235_I}), which received extensive multi-wavelength coverage, displayed significant similarities, among others: significant correlations and short delays between emission at different bands, X-ray spectrum features beyond the absorption expected from our Galaxy (\cite{Madejski:1996}), and the association of flaring episodes with the appearance of superluminal components in VLBI images of the source (\cite{Agudo:2011,juan_0235_I}). 
    
     The similar time span between these episodes, together with older studies of the source (\cite{Raiteri:2005}) that reported flares in previous decades (1992, 1998), have hinted at a pseudo-periodic behavior with a characteristic timescale of 6-8 years (\cite{jorge:2023}).  A similar timescale was suggested by \textcite{Ostorero:2004}, who explained the nearly periodic long-term variability at lower frequencies with a helical model of the jet that precisely matched most of the flares at 8GHz between years 1975-2000 with a period of almost 6 years. The predicted flare in 2004, however, failed to occur, although a period of stronger variability started that peaked in early 2006 and culminated in the historic peak of October 2008.   So far, all attempts at uncovering a significant and clear periodicity in the emission of \object{AO 0235+164} have failed: the emission, even if repeating and presenting striking similarities, is not periodic, which would imply a precise delay and close resemblance between different periods. On the other hand, the flares are a recurrent phenomenon, and the existence of a characteristic timescale for the system related to the apparent delay of 6-8 years between flares cannot be discarded. 
In this regard, there might be other hints of periodic or pseudo-periodic behavior in this source.
    
     In this study, we present data from the most recent flare of \object{AO 0235+164}, in the year 2021, extending the dataset in \textcite{juan_0235_I} by 4 years, from 2019 to 2023. The new episode confirms the relationship between flares at the different spectral ranges, the appearance of superluminal components in VLBI images, and the changing direction of propagation of these components. The timing of this new flare and its detailed characteristics that we present here further strengthen the hypothesis of a pseudo-periodic behavior.
     
    For this work, we have used a standard flat $\Lambda\mathrm{CDM}$ cosmological model with Hubble constant $H_0=\SI{67.66}{km / Mpc}$, as given by \textcite{Planck:2018}.

    \section{Observations}\label{sec:observations}
    
\begin{figure*}
        \centering
        \includegraphics[width=1.0\textwidth]{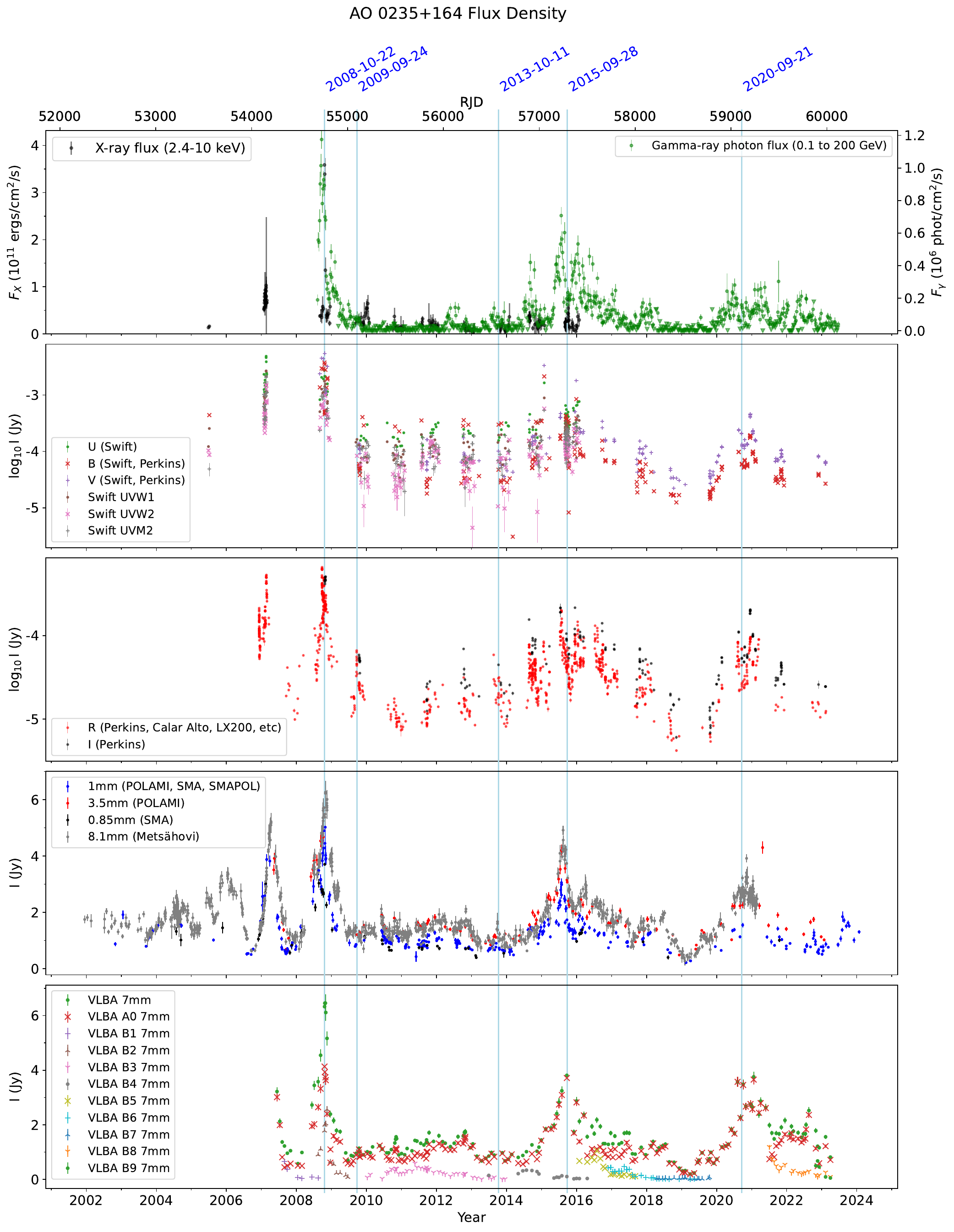}
        \caption{Historical light curves of \object{AO 0235+164} at different wavelengths. The first four vertical lines mark the epochs analyzed in \textcite{juan_0235_I}; the last one corresponds to the epoch analyzed in this work in Sect. \ref{sec:seds}.}
        \label{fig:mwl_flux_all}
    \end{figure*}

    The new dataset presented in this study extends in time by about three years the one in \textcite{juan_0235_I}, and includes %
    7mm (43GHz) VLBA images from the Boston University blazar monitoring program (VLBA-BU-BLAZAR and BEAM-ME programs), reduced both for total flux density and polarization using AIPS (see \cite{Weaver:2022});  %
    single-dish photo-polarimetric data at \si{1}{mm} and \si{3}{mm} from the POLAMI\footnote{{\url{https://polami.iaa.es}}} program at the IRAM 30m Telescope (\cite{polami1}, \cite{polami2}, \cite{polami3}); %
    photometric data at $\SI{1}{mm}$ and $\SI{0.8}{\micro m}$ from the Submillimeter Array (SMA), including %
    photo-polarimetric data at \si{1}{mm} from the SMAPOL program (see Appendix \ref{appendix:SMAPOL} for details);
    8mm observations from the Metsähovi Radio Observatory, and %
    optical data from the Calar Alto (2.2\,m Telescope) under the MAPCAT program and%
    from the Perkins Telescope Observatory (1.8 m Telescope). %
    Gamma-ray data in the \SIrange{0.1}{200}{GeV} range comes from the \textit{Fermi} - Large Area Telescope (LAT).
    The historical light curve shown in Figure \ref{fig:mwl_flux_all} also contains previously published optical data from the Crimea Observatory AZT-8 (0.7 m Telescope) and the St.\ Petersburg State University LX-200 (0.4 m Telescope), as well as ultraviolet data from the \textit{Swift}-UVOT instrument, X-ray data in the \SIrange{2.4}{10}{keV} range from the \textit{RXTE} satellite, and in the \SIrange{0.2}{10}{keV} energy range from \textit{Swift}-XRT (see \cite{juan_0235_I} for details).
    
    We followed the procedure described in \textcite{Blinov:2019} to overcome the $\SI{\pm180}{\degree}$ polarization angle ambiguity in our R-band measurements, minimizing the difference between successive measurements while also taking into account their uncertainty. We also shifted clusters of close observations by an integer multiple of \SI{180}{\degree} to match the angle reported at \si{3}{mm}, enabling visual comparison of the joint evolution of the optical and millimeter range polarization angles, while maintaining the short time evolution intact. Data from the infrared (IR) to the ultraviolet (UV) bands were corrected following the prescription by \textcite{Raiteri:2005} and the updated values by \textcite{Ackermann:2012}. This correction accounts for the local Galactic extinction at $z=0$ and the intervening galaxy ELISA at $z=0.524$, as well as for ELISA's contribution to the observed emission. For details about the correction to the X-ray present in the historical light curve, see \textcite{juan_0235_I}.

    \begin{figure*}
        \centering
        \includegraphics[width=1.0\textwidth]{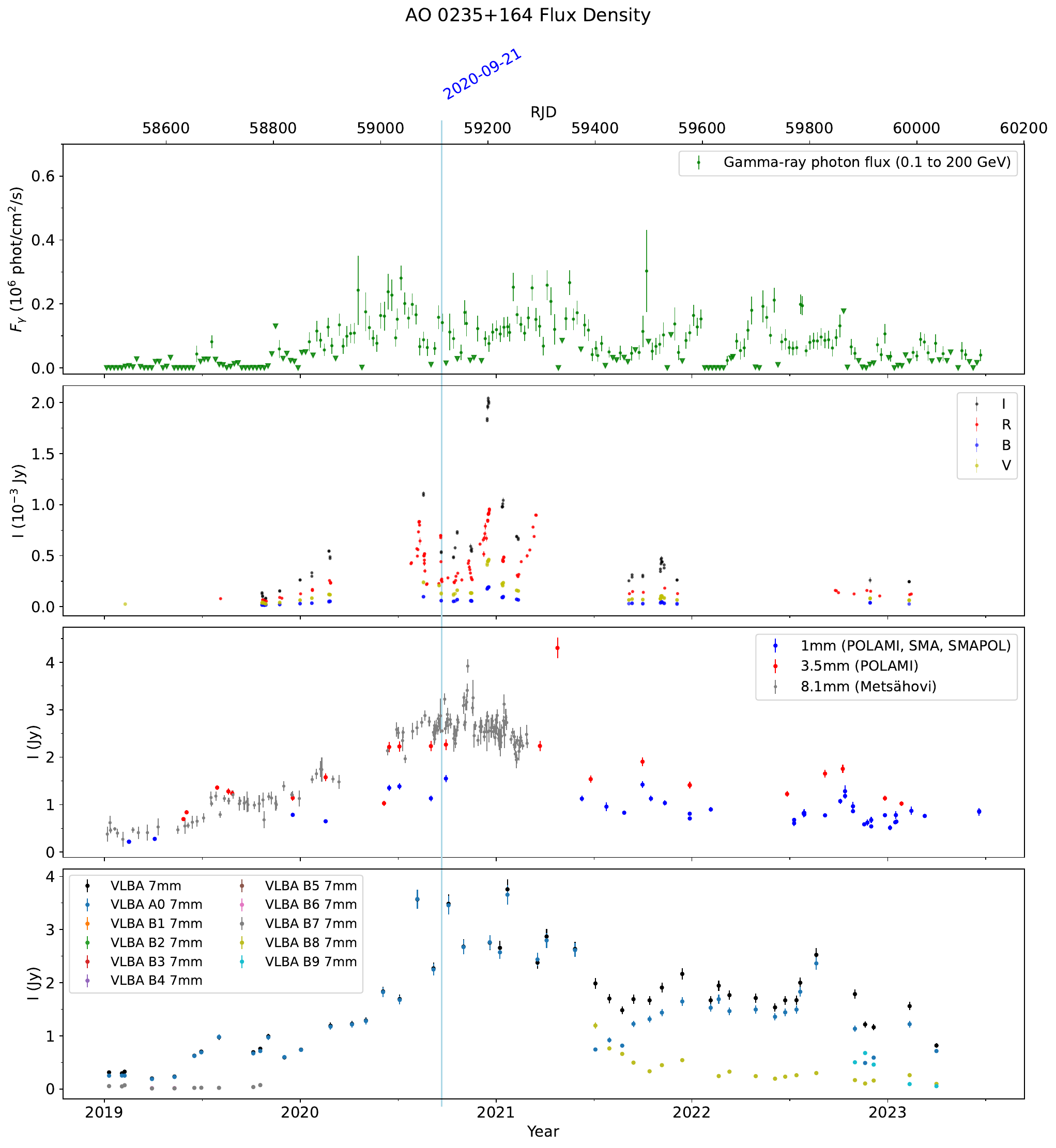}
        \caption{A zoom-in of the flux evolution of \object{AO 0235+164} at different wavelengths during the years 2019-2923. The vertical line corresponds to the last marked date in Fig. \ref{fig:mwl_flux_all}.}
        \label{fig:mwl_flux_2021}
    \end{figure*}
    
    \begin{figure*}
            \centering
            \includegraphics[width=1.0\textwidth]{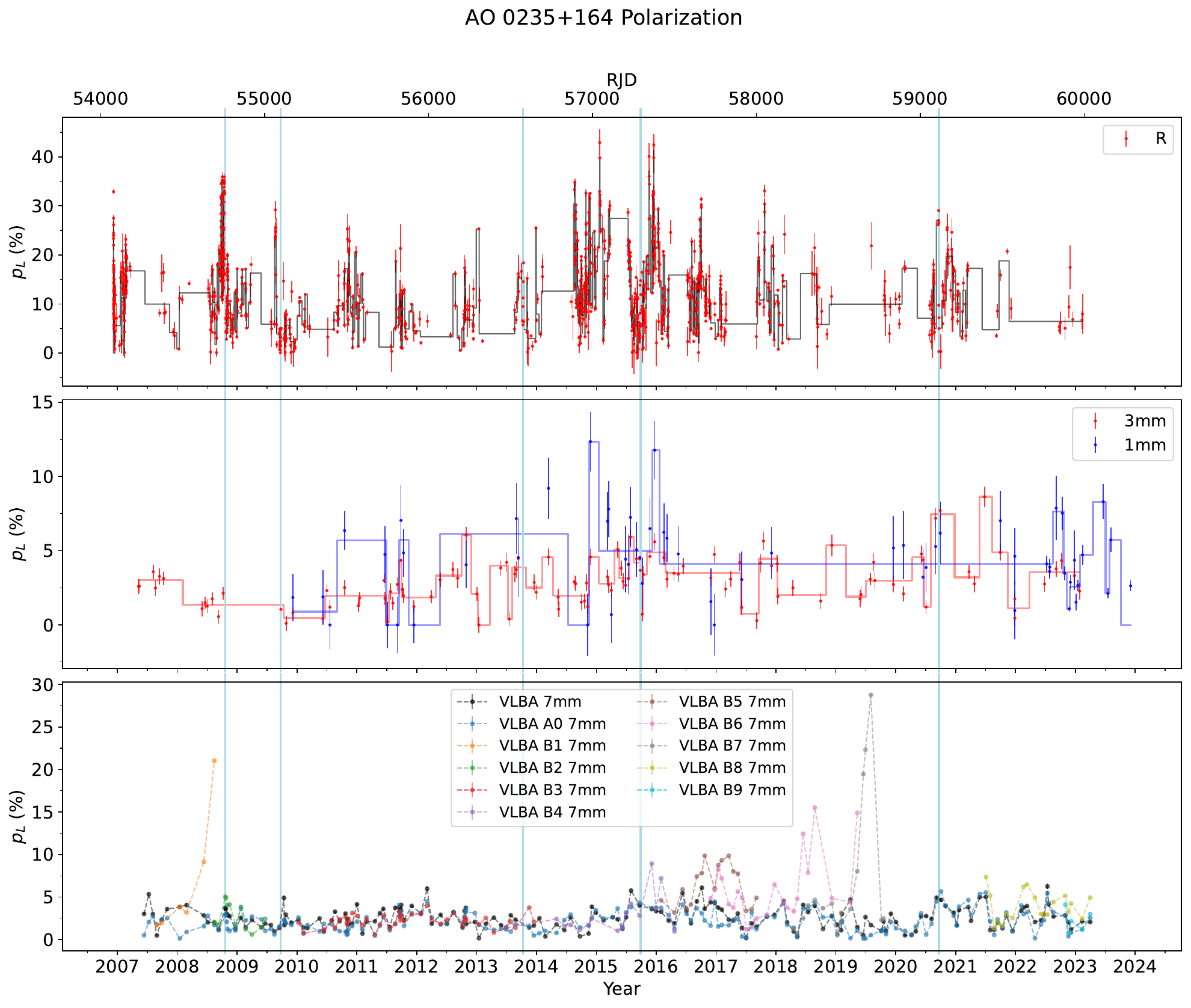}
            \caption{Historical evolution in polarization degree of \object{AO 0235+164}. A Bayesian block representation is shown superimposed for R at 99.9\% confidence and for 1mm and 3mm at 90\% confidence level. Vertical lines corresponds to dates marked in Fig. \ref{fig:mwl_flux_all}.}.
            \label{fig:mwl_pol_I_2021}
    \end{figure*}
    
    \begin{figure*}
        \centering
        \includegraphics[width=1.0\textwidth]{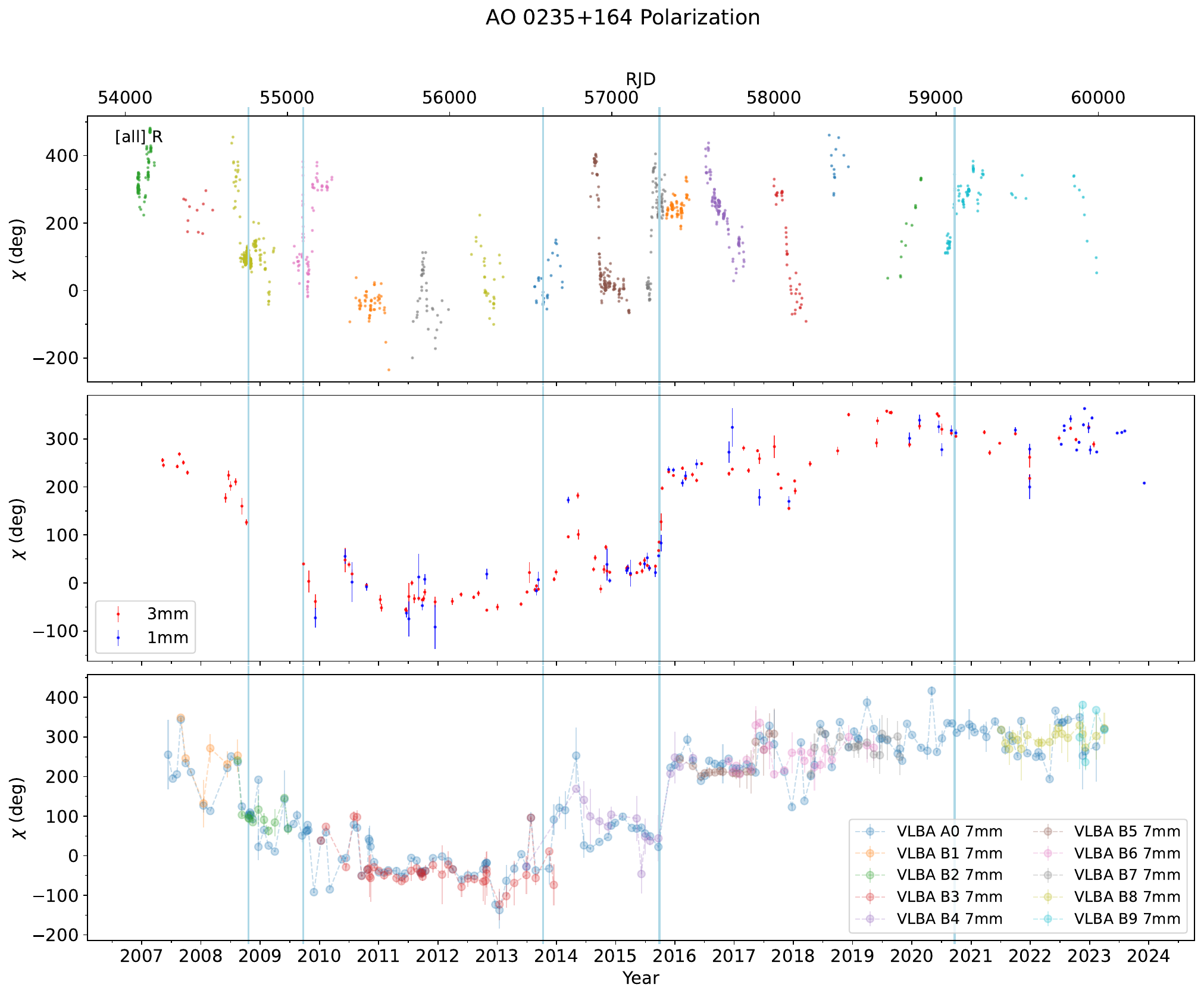}
        \caption{Historical evolution in polarization angle of \object{AO 0235+164}. Vertical lines correspond to dates marked in Fig. \ref{fig:mwl_flux_all}. All points in the first panel correspond to the R band; the different colors denote the clusters that were shifted by $n \times \SI{180}{\degree}$ to follow the evolution of the polarization angle at 3mm, as discussed in Sect. \ref{sec:observations}.}
        \label{fig:mwl_pol_II_2021}
    \end{figure*}

    \begin{figure*}
        \centering
        \includegraphics[width=0.95\textwidth]{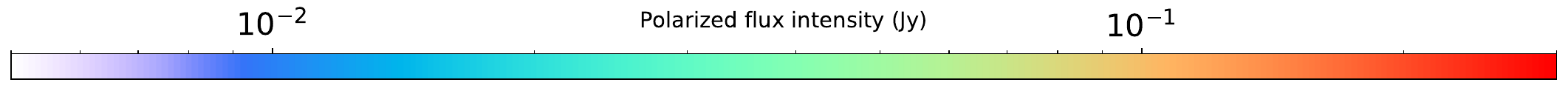}
        \includegraphics[width=0.95\textwidth]{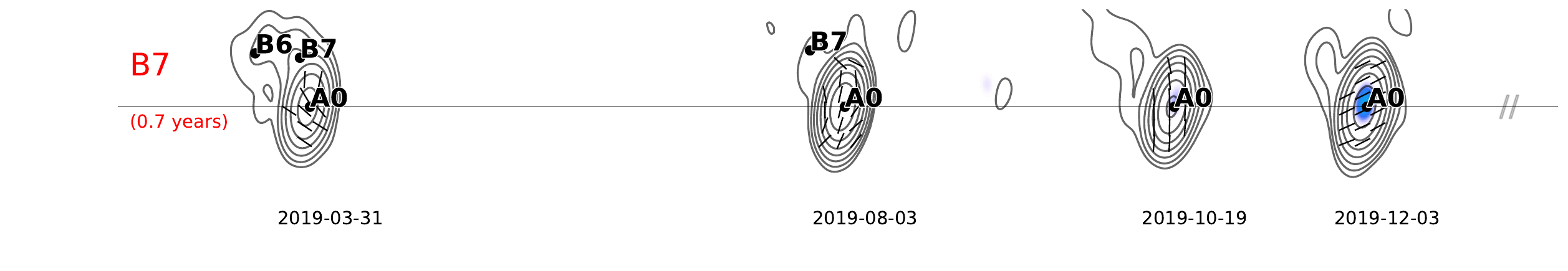}
        \includegraphics[width=0.95\textwidth]{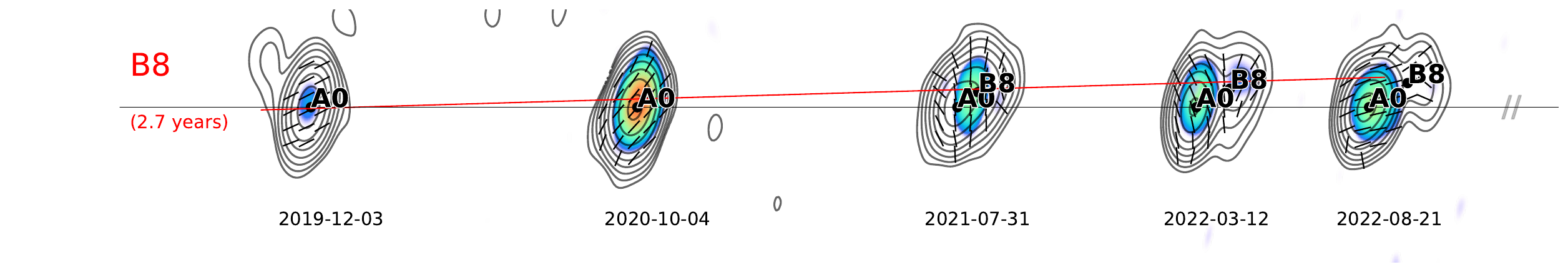}
        \includegraphics[width=0.95\textwidth]{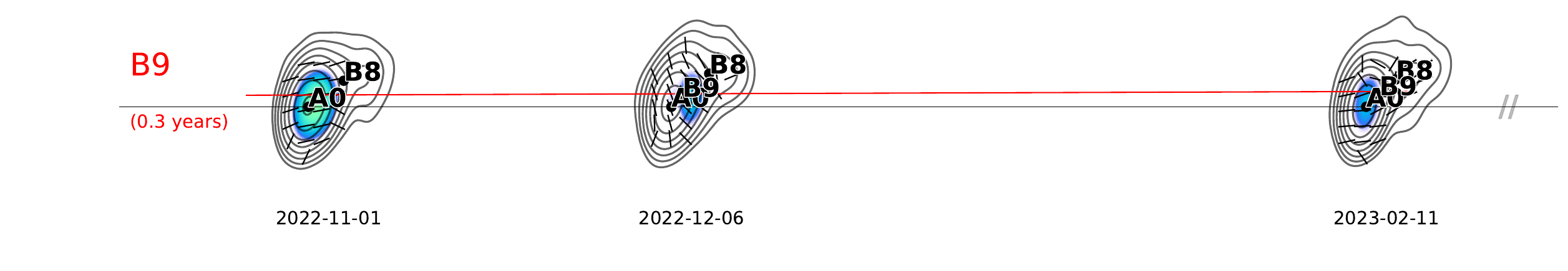}
        \caption{Selected epochs illustrating the evolution of the last 3 identified components, B7, B8 and B9. The figure shows total intensity (contours), polarized flux intensity (color sale) and polarization direction (black line segments). The horizontal black line marks the position of the core A0. The red line in each row is a linear fit to the knot position, present for all except B7 due to its low flux. For each row, the spacing between plots is proportional to the elapsed time, with the total time indicated in brackets. The full temporal evolution is available as an online movie.}
        \label{fig:selected_vlbi_epochs}
    \end{figure*}
    
    \section{Analysis and results}
    
    \subsection{Multi-wavelength flux and polarization behavior}
    
    In Figure \ref{fig:mwl_flux_all}, we present the multi-wavelength light curves consisting of our compiled data in the millimeter, optical, and high-energy ranges from 2008 to 2023. This period includes the three recent flaring episodes of the source in 2008, 2015 and 2021. A detailed view into the last of these flaring episodes can be found in Figure \ref{fig:mwl_flux_2021}. During the flare, emission at all wavelengths experienced an increase, from mm-wave to high-energy $\gamma$-rays, as happened in previous flaring episodes. Compared to previous flares of the source, the 2021 flare was weaker, following a trend that started with the 2015 episode. In agreement with past episodes, the light curve during the 2021 flare shows a multi-peak structure, with sharper variability at higher energies. 
    
    The evolution of both the linear polarization degree and the polarization angle at all wavelengths are shown in Figures \ref{fig:mwl_pol_I_2021} and \ref{fig:mwl_pol_II_2021}, respectively. The polarization degree also increased during the episode, from $p_{L, \mathrm{R}}=\SI{10.0\pm6.0}{\percent}$ and  $p_{L, \mathrm{3mm}}=\SI{3.0\pm1.4}{\percent}$ in the time spanned from 2018 to 2019 to $p_{L, \mathrm{R}}=\SI{12.5\pm6.3}{\percent}$ and  $p_{L, \mathrm{3mm}}=\SI{3.7\pm2.0}{\percent}$ from 2019 to 2023, although this increase was not as dramatic as in the previous flaring episodes (\cite{juan_0235_I}). In contrast to the 2015 episode, the polarization angle  at 3mm remained more stable, with no clear rotations apparent in the available data.

    \subsection{VLBA imaging} \label{sec:vlbi_imaging}
    
     We analyze a total of 49 new Very Long Baseline Array (VLBA) 7mm (43GHz) images of \object{AO 0235+164}, extending the dataset presented in \textcite{juan_0235_I} by about 4 years.
     In our new VLBI images, the most recent flare is accompanied by the ejection of newly emerged components B8 and B9, as can be seen in Figure \ref{fig:selected_vlbi_epochs} for some selected epochs. These components move away from the compact, stationary region present at all epochs known as the ``core''. 
     As in our previous studies, these components were obtained by fitting the reduced VLBA total flux maps to circular Gaussian components in the $(u,v)$ plane using \texttt{Difmap}. After model fitting of the most prominent jet features, we cross-identified them along the different observing epochs. This was done for all new 49 observing epochs. Their resulting flux and polarization evolution is shown in Figures \ref{fig:mwl_pol_I_2021} and \ref{fig:mwl_pol_II_2021}, together with that of the core region (labeled as component A0) and the total integrated emission from the source at 7mm.
    
    The connection between the multi-wavelength (MWL) flare and the ejection of superluminal components in VLBA images at 7mm is confirmed here for the 2021 flare, as was also found for the 2006-2008 flare(s), which was accompanied by the appearances of components B1 and B2, and for the 2015 flare, when components B5, B6 and B7 were ejected. We observe that the increase of brightness of the core (A0) begins almost a year before the new B8 component can be differentiated, although the polarization angle is already aligned with the future direction of B8. This alignment is maintained for most of the lifetime of the component, except for short rotations just after its ejection. This behavior was also reported previously for the superluminal components associated with the 2008 and 2015 flares, and also for components B3 and B4 during the quiescent period in between those two flares (\cite{juan_0235_I}).
    
    Remarkably, the direction of propagation of these new components is very different from that of traveling components identified previously in AO 0235+164 with the VLBA at 7mm. In fact, it has been the case for AO 0235+164 that the direction of ejection has consistently changed from one episode to the next: 
    B1 ($\SI{111\pm3}{\degree}$),  
    B2 ($\SI{-72\pm16}{\degree}$),  
    B3 ($\SI{-73\pm10}{\degree}$) ,  
    B4 ($\SI{153\pm21}{\degree}$),  
    B5 ($\SI{29\pm4}{\degree}$), 
    B6 ($\SI{40\pm10}{\degree}$), 
    B7 ($\SI{70\pm10}{\degree}$), 
    B8 ($\SI{147\pm8}{\degree}$), and 
    B9 ($\SI{144\pm8}{\degree}$). 
        The new flare and its associated traveling components confirm the wobbling of the jet and its narrow viewing angle. A very low viewing angle of the jet is necessary for any reasonably small change in its direction to produce such radical changes in the direction on the sky of propagation of the superluminal components.
    
    \subsection{Kinematic parameters of the VLBI jet components}\label{sec:kinematics}
    
\begin{figure}
        \includegraphics[width=1\linewidth]{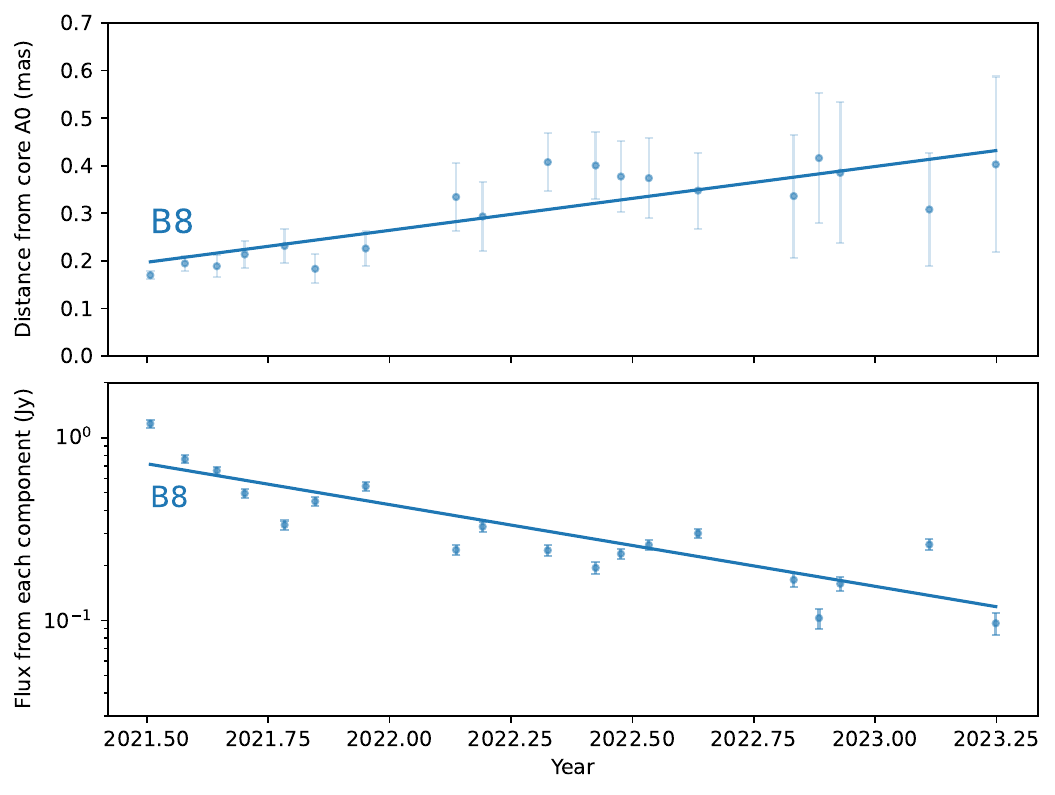}
        \caption{Observed distance from the core (top) and flux density (bottom) for the newly identified component B8, together with the linear weighted fit to the flux knot position and the logarithmic fit to the knot flux used in the computation of the kinematic parameters.}
        \label{fig:kinematic_fits_dist_flux:b8}
    \end{figure}

    We have computed the kinematic parameters of the new B8 component following the procedure described in \textcite{Weaver:2022}, as was done in \textcite{juan_0235_I} for B1 to B6. The procedure involved tracing the identified features in the VLBA images across all new epochs and fitting their positions to a linear function to obtain their speed ($v_{r}$) and time of ejection ($t_0$), as well as fitting their fluxes to a decaying exponential to obtain the timescale of variability ($t_{var}$). This allowed us to compute their Doppler factor $\delta$ and apparent speeds $\beta_{app}$ (\cite{Jorstad:2005}, \cite{Casadio:2015}). From these, the corresponding bulk Lorentz factors $\Gamma$ and viewing angles $\Theta$ could be computed using the usual expressions. 
    As was the case for B7 during the previous flare, the kinematic parameters of B9 could not be correctly estimated due to the low number of observations, and only the time of ejection could be computed for B7 and B9. The fits to the position and flux of B8 can be found in Figure \ref{fig:kinematic_fits_dist_flux:b8}. 
    For B8, we obtained a time of ejection 
    $t_0=\SI{2020.0\pm0.2}{year}$, Doppler factor $\delta_{var}=\SI{25.5\pm1.5}{}$, apparent speed $\beta_{app}=\SI{6.7\pm1.0}{}$, bulk Lorentz factor $\Gamma=\SI{13.6\pm1.7}{}$ and viewing angle  $\Theta=\SI{1.1\pm0.2}{\degree}$. 
    These results are in agreement with those found in \textcite{juan_0235_I}. The Doppler factor of B8, the main ejected component responsible for the 2021 flare, is much lower than that of B2 for the 2008 flare ($\delta=\SI{67.8\pm3.6}{}$) and B5 for the 2015 flare ($\delta=\SI{39.8\pm2}{}$), explaining the relatively diminished luminosity of each flare as a consequence of weaker Doppler boosting.

     \subsection{Change of jet direction}\label{sec:wobbling}
     
     To try to explain the observed wobbling of the jet, we use the derived times of ejection of the superluminal components and their direction of propagation to test for signs of a precessing jet by fitting to an analytical model for the angle of propagation. 
     If we assume that the ejection happens at approximately constant distance from the base of the jet, this location, when projected in the plane of the sky, must trace an ellipse for which the eccentric anomaly $E$ is given by the precessed angle, $E=\omega t$. The eccentricity of this ellipse will be given by the angle between the axis of precession and the observer. In arbitrary units, the semi-axes of this ellipse can be taken to be $a=1$ and $b=\cos{\varphi}$. In the plane of the sky, the center of this ellipse will be displaced with respect to the basis of the jet by a distance $d\sin{\varphi}$, and the major semi-axis will form an arbitrary angle $\psi$ with regard to the x-axis. The polar coordinate of this region orbiting along the ellipse with respect to the basis of the jet is the observed angle of propagation of the superluminal component. If the angle between the precession axis and the observer is high enough compared to the tilt of the jet with regard to the precession axis, the jet will have a definite direction in the sky, and all components will be emitted in the span of some arc.
     When the jet is narrowly pointing towards us, even for small precession angles, the superluminal components will appear to propagate in all directions. This would be the case for \object{AO 0235+164}.
     
\begin{figure}
             \centering
              \includegraphics[width=1\linewidth]{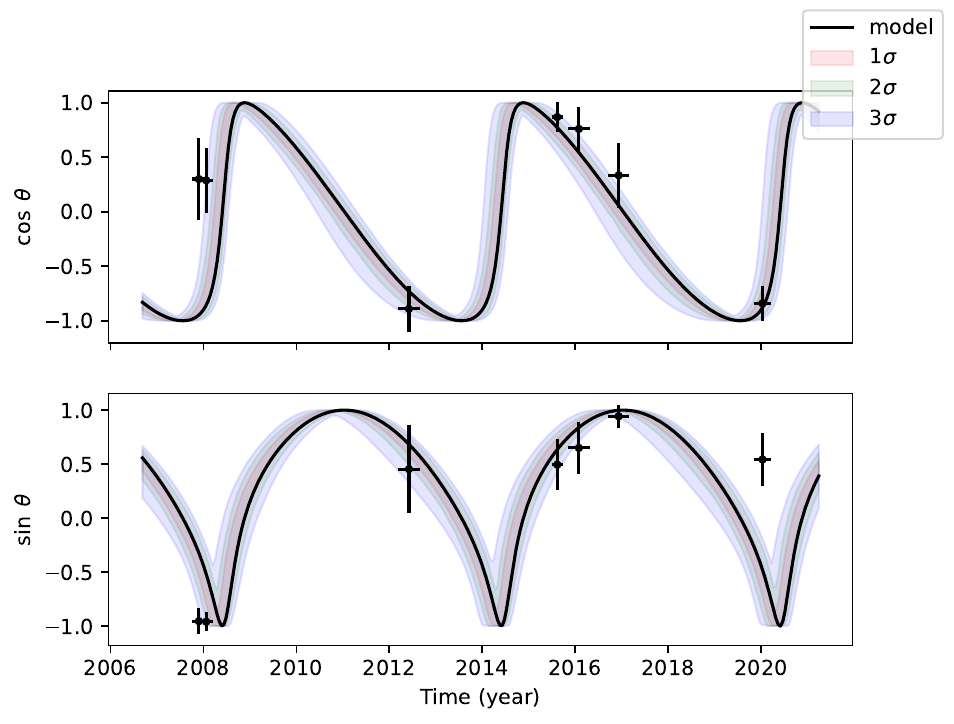}
             \caption{Simultaneous fit to the sine and cosine of the average position angle $\theta$ of the identified VLBI jet features as a function of their computed time of ejection. Most of the points lie within the $3\sigma$ model uncertainty region.  The simultaneous fit avoids the $\pm 180$ uncertainty in the position angle. The corresponding plot for the position angle $\theta$ can be found in Fig. \ref{fig:prec_fit_theta}.}
             \label{fig:prec_fit}
         \end{figure}

\begin{figure}
            \centering
             \includegraphics[width=0.9\linewidth]{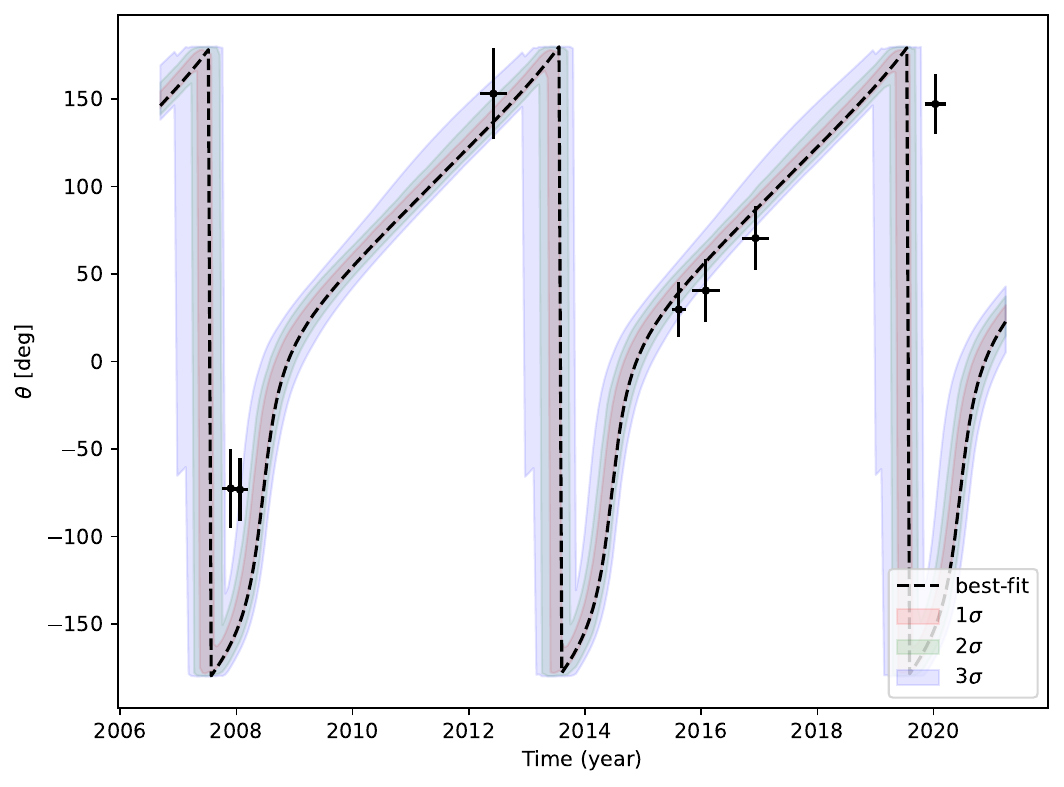}
            \caption{Observed ejection angles of the identified VLBI components B2-B8 with respect their computed time of ejection, together with the best-fit model. The shaded areas represent the $1\sigma$ (68.27\%), $2\sigma$ (95.45\%), $3\sigma$ (99.73\%) uncertainties of the model.}
            \label{fig:prec_fit_theta}
        \end{figure}

    Figure \ref{fig:prec_fit} shows the result of simultaneously fitting $(t_0,\cos{\theta})$ and  $(t_0,\sin{\theta})$, where $\theta$ is the propagation angle of a component. This fit was preferred to directly fitting the angle $\theta$ because it removes any ambiguity in the position angle. The uncertainties in the parameters were computed using a Monte Carlo approach. Only components B2 to B8 were used for the fit, on the basis of selecting only those with a significant number of epochs to compensate for the high uncertainties in the positions of the fitted Gaussian features mentioned in Sect.\ \ref{sec:vlbi_imaging}.  We have examined the impact of considering all components in the fit and concluded that the optimal fit values were comparable, despite a significant increase in the model uncertainties.
    The resulting best-fit model has a period of $T=\SI{6.0\pm0.1}{years}$, which is within the range of periods proposed in the literature.
    The obtained eccentricity of the ellipse gives an angle of $\phi=\SI{0.11}{\degree}$ for the hypothetical precession axis with regard to the observer, although this value is not well constrained by our fit.
    Because of the low number of points and  the significant uncertainties in the times of ejection of jet features, usual fit statistics such as the p-value are not suitable to reaffirm or reject our hypothesis, and therefore do not allow us to claim precession in \object{AO 0235+164}. %
    Nevertheless, we find it interesting to examine Figures \ref{fig:prec_fit} and \ref{fig:prec_fit_theta}, where it can be seen that most of the points fall inside the $3\sigma$ region when accounting for their uncertainties. We propose that the model might be used to affirm or discard the hypothesis of precession when more data become available in subsequent decades. We also emphasize that the value for the period found in this analysis, derived from the propagation angle of superluminal components,  is independent of the periods suggested in the literature (\cite{jorge:2023},  \cite{Raiteri:2005} \cite{Ostorero:2004}), which are derived from analysis of variations in the light curves. They nevertheless agree within their uncertainties. 
    The small value obtained for the possible precession axis relative to the line of sight, together with the viewing angles obtained in the kinematic analysis (Sect.\ \ref{sec:kinematics}), is also in agreement with the observed behavior of the jet, which ejects components in completely different directions.

    \subsection{Correlations across the spectrum} \label{sec:correlations}
    
\begin{figure*}
        \includegraphics[width=1\textwidth]{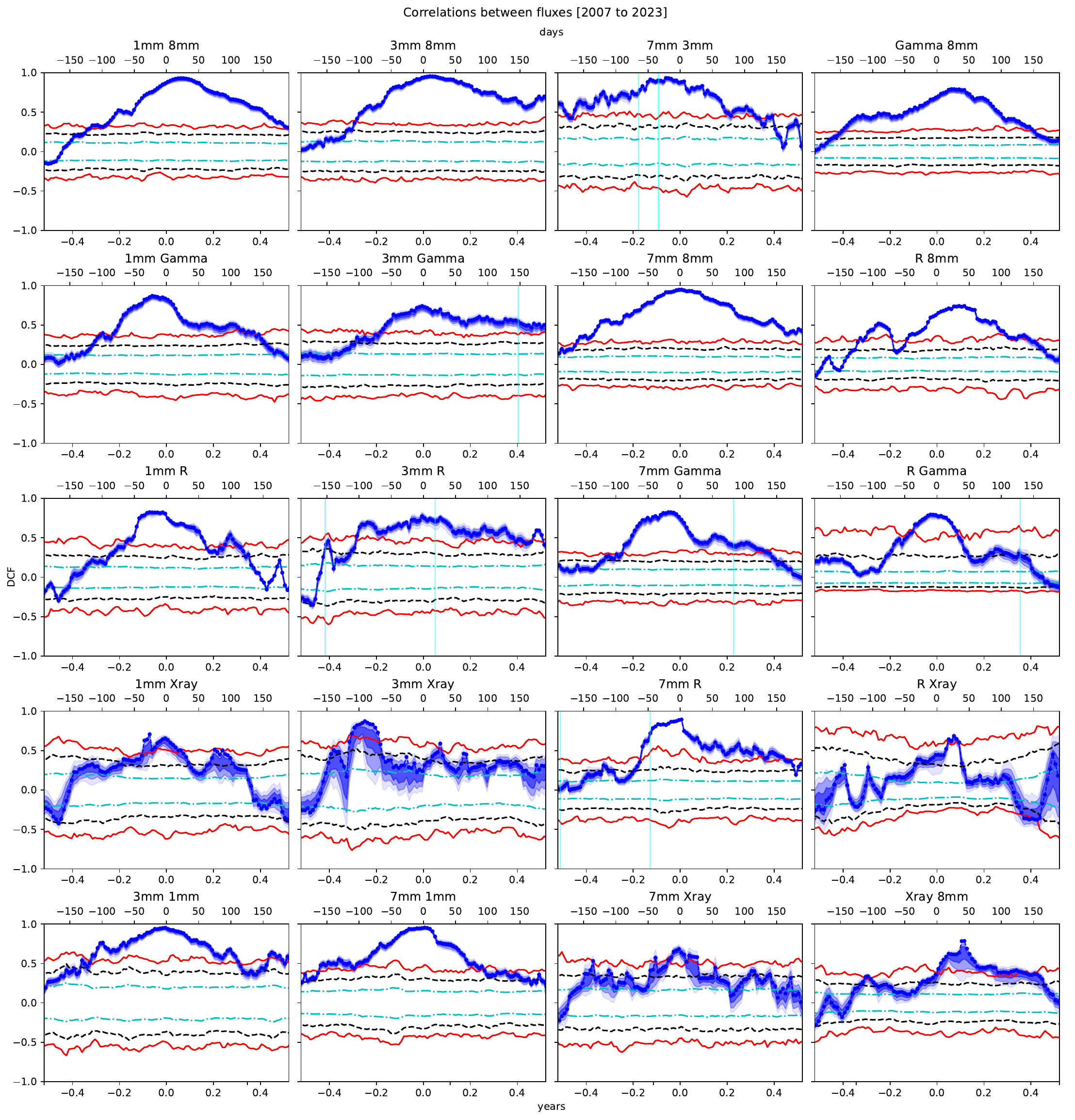}
        \includegraphics[width=0.285\textwidth]{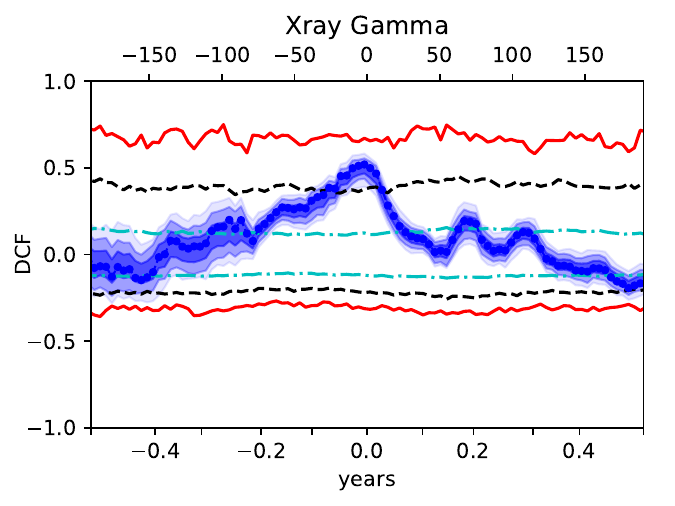}
        \caption{Correlations between fluxes at all wavelengths. Horizontal lines represent the $1\sigma$,  $2\sigma$, and $3\sigma$ significance levels, and were computed using a Monte-Carlo approach with $N=2000$ synthetic light curves. We find clear and significant ($>3\sigma$) correlations near zero between all bands except for X-ray.}
        \label{fig:corr_I}
    \end{figure*}

    We have computed the correlations between the light curves at different wavelengths using \texttt{MUTIS}%
    \footnote{MUltiwavelength TIme Series.  A Python package for the analysis of correlations of light curves and their statistical significance. \url{https://github.com/IAA-CSIC/MUTIS}}. %
    We used the normalized discrete correlation function (DCF) proposed by \textcite{Welsh:1999}, which applies normalization and binning, making it suitable for our irregularly sampled signals. A uniform bin size of $\SI{20}{days}$ was used, a value that was chosen to allow for enough bin statistics without smoothing the correlations too much. To validate our choice, the same results were derived using binning sizes from $\SI{10}{days}$ to $\SI{30}{days}$, confirming the consistency of our results.
    The significance of the correlations was estimated using a Monte-Carlo approach, generating $N=2000$ synthetic light curves for each signal. Randomization of the Fourier transform was used for mm-wavelengths, generating light curves with similar statistical properties and power-spectral density (PSD). For optical and $\gamma$-ray data we modeled the signals as Orstein-Uhlenbeck stochastic processes (\cite{Bonnoli:2004}), which better reproduces the qualitative shape of these signals. The uncertainties of the correlations were estimated using the uncertainties of the signals, again with a Monte-Carlo approach. 
    We found high and significant correlations between emission from all bands except X-rays (Fig. \ref{fig:corr_I}), for which correlation was generally lower and found only at $2\sigma$ with some bands. 
     This is in agreement with our previous results (\cite{juan_0235_I}) that found decreased correlation for the X-ray band and attributed it to a different emission mechanism located at a different region.
      Nonetheless, the achieved significance of those correlations involving X-rays are generally higher than those previously found, thanks to the improved dataset. This is especially the case for the correlation between X-rays and 1mm, where no significant correlation was previously found.
    This decreased correlation for the X-ray is expected also from the spectral energy distribution modeling presented in Sect.\ \ref{sec:seds}, and also that of \textcite{juan_0235_I}, where the bulk Compton emission dominates the X-rays in its high state, while in its low state it is dominated by the same region responsible for emission at other bands.
    Regarding correlations restricted only to the last episode (2019-2023), the new flare is too weak and data too sparse to produce any meaningful correlations, which are dominated by noise, so they have been omitted here.
    
\begin{figure}
        \centering
        \includegraphics[width=1\linewidth]{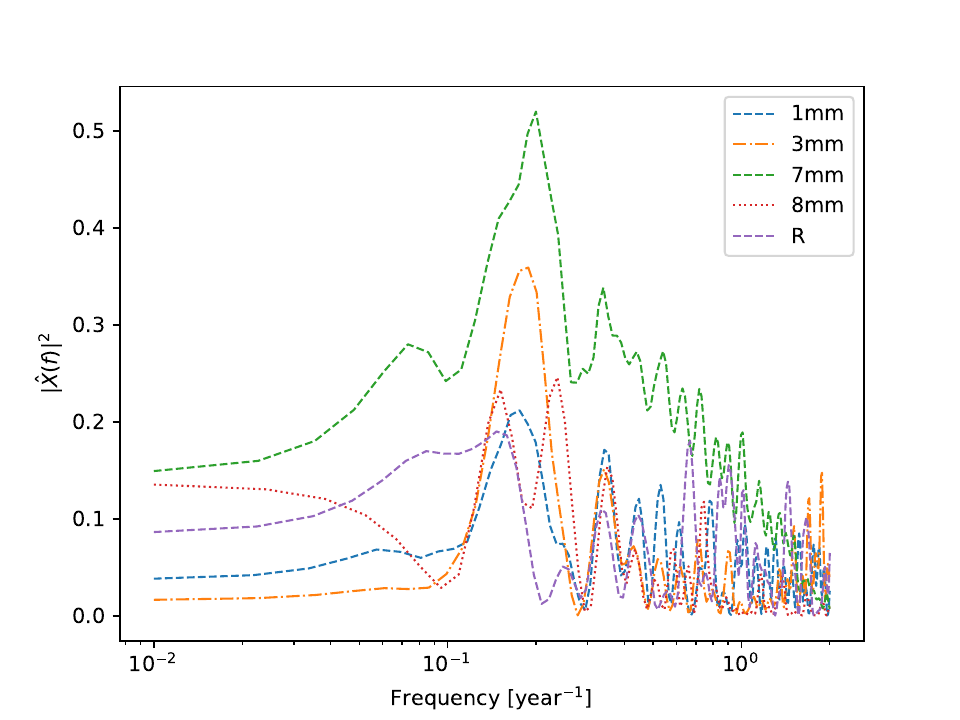}
        \caption{Normalized PSD computed using the Lomb-Scargle periodogram for the 1mm, 3mm, 7mm, 8mm and R light curves, showing peak frequencies corresponding to characteristic timescales of 5.7, 5.4, 5.1, 4.2 and 6.8 years respectively. These timescales are mostly in agreement with the 5-8 year timescale found in previous works. The false alarm probability in all cases is close to zero ($\ll \SI{0.1}{\percent}$), meaning that there is a very low probability that such a peak would be caused by a purely noise signal.}
        \label{fig:LS-all}
    \end{figure}

    Analysis of the power spectrum density (PSD) of the signals was performed using the Lomb-Scargle periodogram (\cite{Lomb:1976, Scargle:1982}), suitable for our unevenly sampled light curves. The resulting power spectra locate the peak frequencies of the 1mm, 3mm, 7mm, 8mm and R-band light curves at equivalent timescales of 5.7, 5.4, 5.1, 4.2 and 6.8 years, respectively (Fig. \ref{fig:LS-all}). Computed false-alarm probabilities are close to zero in all cases ($\ll\SI{0.1}{\percent}$). The interpretation of this probability is subtle (\cite{VanderPlas:2018}), but instead hints at a low probability of a purely stochastic process, since it represents the probability of a purely noise signal producing a peak higher than ours. The derived timescales agree with those suggested in the literature (\cite{jorge:2023, Raiteri:2005, Ostorero:2004}) and with the one independently obtained in Sect.\ \ref{sec:wobbling}.
    
    \subsection{Spectral energy distribution} \label{sec:seds}
    
      \begin{figure}[htbp!]
        \centering
        \includegraphics[width=1\linewidth]{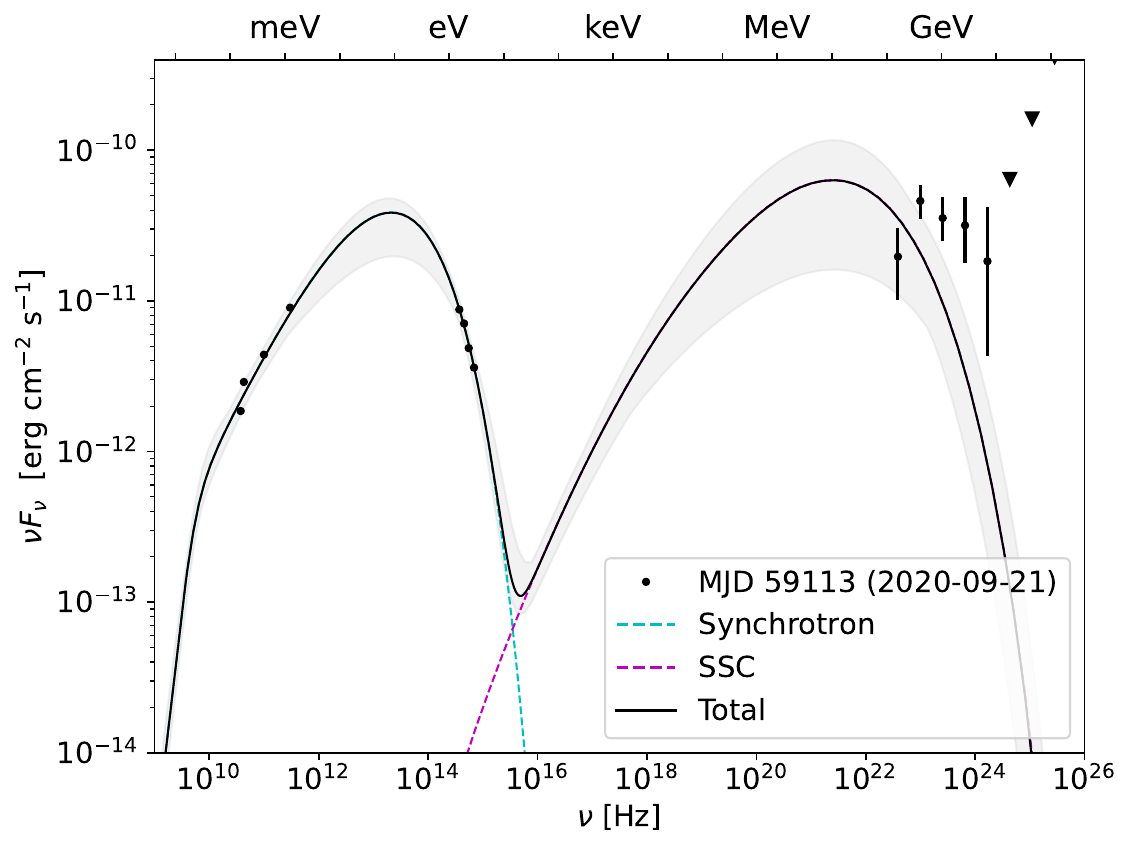}
        \caption{Spectral energy distribution for 2023-09-21, together with the best-fit of the 1-zone SSC model discussed in Section \ref{sec:seds}. The gray area represents the $3\sigma$ uncertainty region. The downward-pointing triangles represent upper limits.}
        \label{fig:sed_model_ssc_2020-09-21}
    \end{figure}
    
    \begin{figure}[htbp!]
        \centering
        \includegraphics[width=1\linewidth]{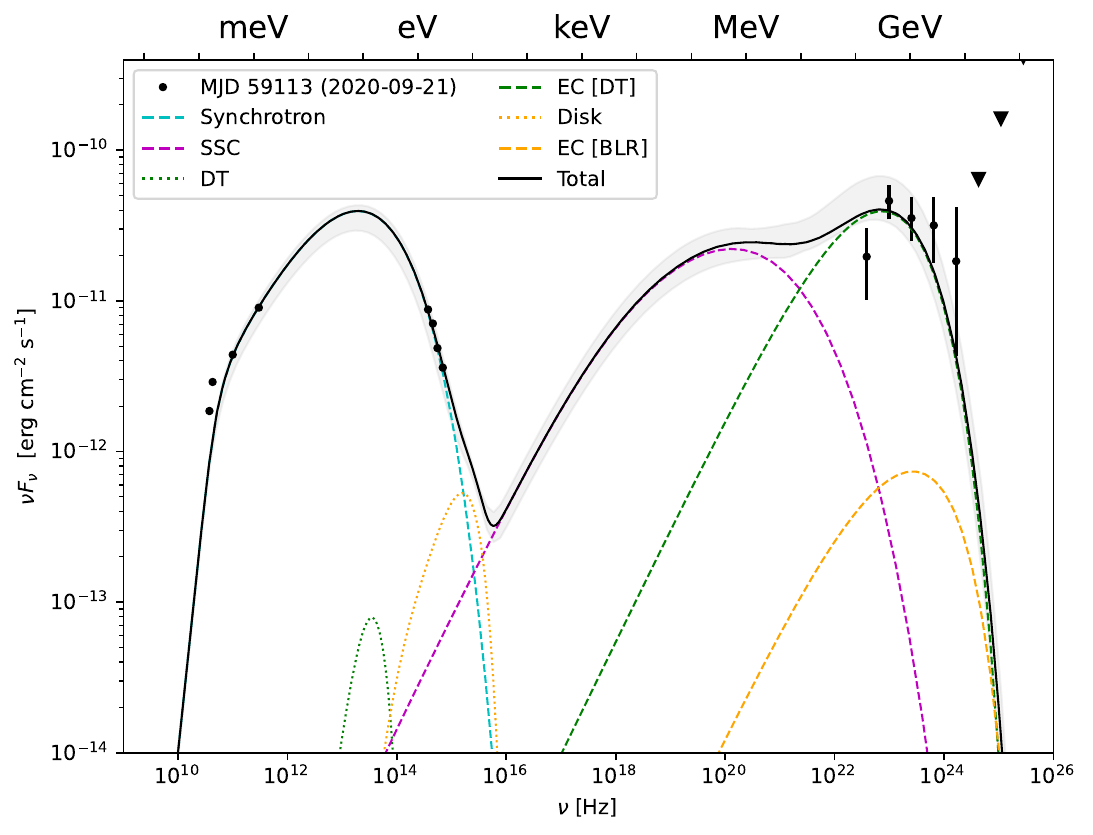}
        \caption{Spectral energy distribution for 2023-09-21, together with the best-fit of the SSC+EC model with freely varying $\Gamma$ and $\theta$ as discussed in Section \ref{sec:seds}. The gray area represents the $3\sigma$ uncertainty region. The downward-pointing triangles represent upper-limits.}
        \label{fig:sed_model_ssc+ec_m1_2020-09-21}
    \end{figure}

    \begin{figure}[htbp!]
        \centering
        \includegraphics[width=1\linewidth]{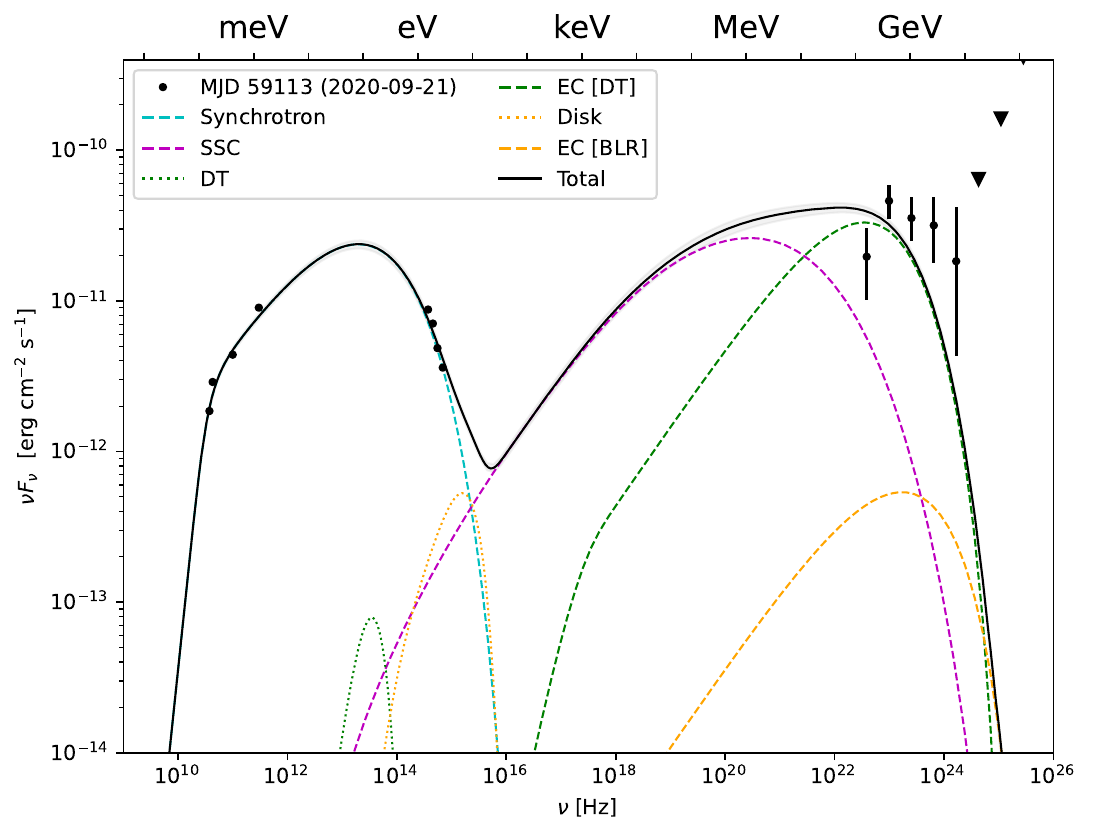}
        \caption{Spectral energy distribution for 2023-09-21, together with the best-fit of the SSC+EC model with fixed $\Gamma$ and $\theta$ as discussed in Section \ref{sec:seds}. The gray area represents the $3\sigma$ uncertainty region. The downward-pointing triangles represent upper-limits.}
        \label{fig:sed_model_ssc+ec_m2_2020-09-21}
    \end{figure}
        
    Unlike for previous flares, no \textit{Swift} XRT or UVOT data are available during the flaring period in 2021. The night with the broadest MWL coverage (MJD 59113, 2020-10-20) was selected to perform a spectral energy model of the source.
    
    We have modeled the emission using the JetSeT framework (\cite{Tramacere:2020, Tramacere:2011, Tramacere:2009}) using both a single-zone synchrotron self-Compton (SSC) scenario and an SSC plus external Compton (EC) scenario. The common physical setup consists of an spherical emitting region formed by a population of relativistic electrons of radius $R$, at a distance $R_H$ from the central black hole, that moves at a small angle $\theta$ to the line of sight with a bulk Lorentz factor $\Gamma$.  Synchrotron radiation is emitted through interaction of the relativistic electrons with the jet magnetic field $B$. These same synchrotron photons and electrons interact by inverse Compton scattering to produce high-energy photons (SSC). %
    In the EC scenario, additional radiation is produced from inverse Compton scattering of photons coming from a disk torus (Disk) and broad line region (BLR) surrounding the BH.
    The electron energy distribution is assumed to be well described by a power law with a cutoff (PLC):
    \begin{equation}
        n(\gamma) = N \gamma^{-p} \exp{\frac{\gamma}{\gamma_{\rm cutoff}}}
        ~,~
           \gamma_{min}\leq \gamma \leq \gamma_{max}
        ~.
        \label{eq:sed_plc}
    \end{equation}
   where $p$ is the spectral index and $\gamma$ is the electron Lorentz factor. A broken power law (BKN) was also attempted, but the resulting fit was systematically worse, consistent with the results of \textcite{juan_0235_I}, and therefore we only show the model with the PLC distribution. 
   
   In the single-zone SSC scenario, it was found that the region of emission was best described as having a radius 
   $R = {2.77}_{ -0.17}^{ +0.18} \times 10^{18} ~ \mathrm{cm}	$, 
   situated at a distance 
   $R_{H} = \SI{9.3e+19}{cm}$ with a bulk Lorentz factor 
   $\Gamma = \num{14.0}_{   -1.2}^{   +0.9} $, 
   and a viewing angle of
   $\theta        =    \SI{0.89}{\degree} {}_{   -0.16}^{   +0.12} $, 
   while the electron distribution was best modeled as having a spectral index
   $p                  =    {1.47}_{   -0.07}^{  +0.08} $,
   and minimum and maximum Lorentz factors of 
   $\gamma_{min} = {105}_{   -15}^{   +15}  $,
   $\gamma_{max} = {1.72}_{   -0.3}^{   +0.3} \times 10^{+5} $,
   and a cutoff of
   $\gamma_{cut} = {7.7}_{   -0.5}^{   +0.5} \times 10^{+3} $.
   The best-fit value for the magnetic field was 
   $B = 3.6_{  -0.4}^{  +0.4} \times 10^{-3}~ \unit{G} $.
   All values reported are best-fit values, with $1\sigma$ asymmetric errors computed using Markov Chain Monte Carlo (MCMC) approach. The result has been represented in Fig. \ref{fig:sed_model_ssc_2020-09-21}.
   The obtained parameters are in agreement with those obtained by \textcite{juan_0235_I}, and the resulting bulk Lorentz Factor, viewing angle and Doppler factor ($\delta=\SI{25}{}$) agree precisely with those obtained in the kinematic analysis of section \ref{sec:kinematics} within their uncertainties.
    
   In the EC scenario, disk, DT and BLR parameters were frozen. 
   The disk was assumed to have luminosity
   $L_\text{Disk} = \SI{5e45}{erg s^{-1}}$, accretion efficiency $\eta=\num{0.08}$, and inner and outer radii $R_\text{Disk, in}=\SI{3}{R_s}$ and $R_\text{Disk,out}=\SI{300}{R_s}$, respectively. 
   The dusty torus temperature was fixed to $T_\text{DT}=\SI{830}{K}$, its radius %
   determined by the phenomenological relation 
   $R_\text{DT} = \num{2e19} L_\text{Disk,46}^{1/2} \unit{cm}$ (\cite{Cleary:2007}),
   with reprocessing factor $\tau_{DT}=0.1$.
   The BLR was modeled as a thin spherical shell with an internal radius
   as provided by the phenomenological relation $R_\text{BLR,in}=\num{3e17} L_\text{Disk,46}^{1/2} \unit{cm}$ (\cite{Kaspi:2007}) and its outer radius was assumed to be  $R_\text{BLR,out}=1.1 R_\text{BLR,in}$, with a coverage factor $\tau_\text{BLR}=\num{0.1}$.
   The mass of the black hole was set to $M_\text{BH}=\SI{5e8}{M_{\odot}}$.
   
   Two alternative SSC+EC models were produced, one allowing $\Gamma$ and $\theta$ to vary freely, and another fixing them to the values obtained from the kinematic analysis.
   In the former case, it was found that the emitting region was best described by best-fit values
   $\theta = {1.68}_{-0.04}^{+0.08}~\si{\degree}$,
   $\Gamma = \num{50}_{-5.75}^{+1.55}$,
   $B = \num{5.3}_{-0.3}^{+0.5} \times 10^{-2} ~ \si{G}$,
   $N = \num{113}_{-10}^{+10} \si{cm^{-3}}$,
   $\gamma_{min} = \num{1.12}_{-0.06}^{+0.12}$,
   $\gamma_{max} = 6.48_{-0.6}^{+0.3} \times 10^{5}$,
   $\gamma_{cut} = 1.83_{-0.1}^{+0.9} \times 10^{3}$,
   $p = 1.50_{-0.03}^{+0.11}$.
   In the latter, fixing $\Gamma=\num{13.6}$ and $\theta=\SI{0.9}{\degree}$, we obtained 
   $B = \num{2.59}_{-0.05}^{+0.05} \times 10^{-2} ~ \si{G}$,
   $N = \num{88.4}_{-0.2}^{+0.2} \si{cm^{-3}}$,
   $\gamma_{min} = \num{9.17}_{-0.02}^{+0.02}$,
   $\gamma_{max} = 3.48_{-0.07}^{+0.08} \times 10^{4}$,
   $\gamma_{cut} = 4.27_{-0.08}^{+0.08} \times 10^{3}$,
   $p = 1.94_{-0.004}^{+0.004}$.
   The higher bulk Lorentz factor in the first cause with respect to that measured at the VLBI case might be attributed to deceleration.
   The results have been represented in Figs. \ref{fig:sed_model_ssc+ec_m1_2020-09-21} and \ref{fig:sed_model_ssc+ec_m2_2020-09-21}.
   In any case, for all models the resulting Doppler factors are lower than that obtained for the flaring epochs in the 2008 ($\delta=\SI{37}{}$) and 2015 ($\delta=\SI{32}{}$) episodes (\cite{juan_0235_I}), which consistently explains the lower apparent luminosities of the successive flares as caused, at least partially, by relativistic effects.

   \section{Discussion and conclusions}
    
    We have presented new and updated data from the blazar \object{AO 0235+165}, extending previous works to cover its most recent flaring episode that peaked in 2021.
    
    The new flare is again associated with the appearance of two new components in 7mm VLBA images, B8 and B9. The behavior of B9 and B8 is compatible with that of trailing components (\cite{Agudo:2001}), as was the case with B6 and B5 during the 2015 episode. This can be interpreted in the context of a shock-in-jet model, in agreement with the alignment of the polarization angle in the direction of the jet axis that begins during the core re-brightening phase (Fig. \ref{fig:selected_vlbi_epochs}).
    The two newly identified components B8 and B9 propagate in a different direction compared to previous components, confirming the wobbling of the jet.
    
    We have proposed a purely geometrical model that aims to explain the observed changes in the direction of ejection of the VLBI components as the result of a precessing jet. We have found that the observed position angles and calculated times of ejection are mostly compatible with a jet precessing with a period of $\SI{6}{years}$. This value is independently obtained but compatible with those found in the existing literature.
    Although precession is a strictly periodic phenomenon and no exact periodicity is found in the MWL light curves of \object{AO 0235+164}, it is important to keep in mind that the model in Sect.\ \ref{sec:wobbling} relates only to the position angle of the ejected components. %
    However the periodicity found in this model must indeed have an origin that could justify this timescale of variability -not periodicity- of $6-8$ years found in the light curves in previous works (\cite{Raiteri:2005}, \cite{jorge:2023})%
    , and in Sect.\ \ref{sec:correlations}. 
    The absence of a strict periodicity in the flux evolution of the source is to be expected due to the fact that jet emission is a complex process that can be affected by many causes (jet angle, speed, magnetic field, matter accretion, available energy, to name only a few) and is inherently stochastic. %
    However, the possible origin of wobbling, or jet precession, are more restricted, the most common causes being a binary system (\cite{Abraham:2018}) or an off-axis accretion disk (Lense-Thirring effect), and this exact periodicity can be distortedly reflected in light curves. A precessing jet with a periodicity around $\sim\SI{6}{years}$ motivates the appearance of the pseudo-periodic timescale found in the light curves by previous works.
        
   Modeling of the spectral energy distribution reveals that the emission process is similar to that of previous epochs, both flaring and quiescent (\cite{juan_0235_I}), with the difference in Doppler factor explaining at least partially the flux variability. This is the case for all bands, but cannot be confirmed in X-rays due to the absence of data during the 2021 flaring episode. However, in previous works it was found that emission in X-rays was caused at least partially by different, uncorrelated mechanisms involving a different emitting region, and that this region was responsible for the bulk of emission when the X-ray emission was in its high state (\cite{Ackermann:2012, juan_0235_I}).
   Therefore, more complex models are necessary to fully explain the MWL emission. Moreover, even these aforementioned models fail to include hadronic processes, whereas there is increasing evidence of neutrino emission in blazars (\cite{icecube:2018}).
   
   The re-brightening of the core suggests the existence of a standing shock, with the ejected components that accompany each flaring episode interpreted as trailing components (\cite{Agudo:2001}). Such stationary shocks can be explained by bends in the jet, which are expected in wobbling jet scenarios, although such bends need not be caused by rotations of the jet nozzle, but can also be the result of dynamical processes. In addition, stationary shocks can be explained by the interaction of the jet with the external medium (\cite{Gomez:1997}).
   In any case, the recurrence of the flaring episodes, the wobbling of the jet, and the modeling of the spectral energy distribution suggest the existence of a characteristic timescale of a periodic and geometric origin that must be well characterized to achieve a full understanding of the mechanisms of emission of  \object{AO 0235+164}.
        
    \section*{Acknowledgments}
The IAA-CSIC team acknowledges financial support from the Spanish "Ministerio de Ciencia e Innovación" (MCIN/AEI/ 10.13039/501100011033) through the Center of Excellence Severo Ochoa award for the Instituto de Astrofísica de Andalucía-CSIC (CEX2021-001131-S), and through grants PID2019-107847RB-C44 and PID2022-139117NB-C44.
This research has made use of the NASA/IPAC Extragalactic Database (NED),
which is operated by the Jet Propulsion Laboratory, California Institute of Technology,
under contract with the National Aeronautics and Space Administration.
IRAM is supported by INSU/CNRS (France), MPG (Germany) and IGN (Spain).
The VLBA is an instrument of the National Radio Astronomy Observatory, USA. The National Radio Astronomy Observatory is a facility of the National Science Foundation operated under cooperative agreement by Associated Universities, Inc.
This study was based (in part) on observations conducted using the 1.8 m Perkins Telescope Observatory (PTO) in Arizona (USA), which is owned and operated by Boston University.
The BU group was supported in part by U.S. National Science Foundation grant AST-2108622, and NASA Fermi GI grants 80NSSC23K1507, 80NSSC22K1571 and 80NSSC23K1508.
The Submillimeter Array is a joint project between the Smithsonian Astrophysical Observatory and the Academia Sinica Institute of Astronomy and Astrophysics and is funded by the Smithsonian Institution and the Academia Sinica. We recognize that Maunakea is a culturally important site for the indigenous Hawaiian people; we are privileged to study the cosmos from its summit.
        
\nocite{*}
\bibliography{includes/citations} 
    
\begin{appendix}    

\section{SMAPOL Observations} \label{appendix:SMAPOL}
The Submillimeter Array (SMA, \cite{Ho2004}) was used to obtain polarimetric millimeter radio measurements at 1.3~mm (230~GHz) within the framework of the SMAPOL (\textbf{S}MA \textbf{M}onitoring of \textbf{A}GNs with \textbf{POL}arization) program. SMAPOL follows the polarization evolution of forty $\gamma$-ray bright blazars, including AO\,0235+164, on a bi-weekly cadence, as well as other sources in a target-of-opportunity (ToO) mode. The observations reported here were conducted between July 2022 and December 2023.

The SMA observations use two orthogonally polarized receivers, tuned to the same frequency range in full polarization mode, and use the SWARM correlator (\cite{Primiani2016}). These receivers are inherently linearly polarized but are converted to circular using the quarter-wave plates of the SMA polarimeter (\cite{Marrone2008}). The lower sideband (LSB) and upper sideband (USB) covered 209-221 and \num{229}–\si{241}{GHz}, respectively. Each sideband was divided into six chunks, with a bandwidth of \si{2}{GHz}, and a fixed channel width of \si{140}{kHz}. The SMA data were calibrated with the MIR software package \footnote{\href{https://lweb.cfa.harvard.edu/~cqi/mircook.html}{https://lweb.cfa.harvard.edu/~cqi/mircook.html}}. Instrumental polarization leakage was calibrated independently for USB and LSB using the MIRIAD task \texttt{gpcal} (\cite{Sault1995}) and removed from the data.  The polarized intensity, position angle, and polarization percentage were derived from the Stokes I, Q, and U visibilities.

\object{AO 0235+164} was observed 14 times within the above period on June 1, 4 and 16 with integration times between 2.4 and 15 minutes. The total flux density, linear polarization degree and polarization angle results are given in the table attached. \object{MWC 349 A}, \object{Callisto}, \object{Uranus}, \object{Neptune} and \object{Ceres} were used for the total flux calibration according to their visibility, and the calibrator \object{3C 286}, which has a high linear polarization degree and stable polarization angle, was observed regularly as a cross-check of the polarization calibration.
\end{appendix}

\end{document}